\documentclass[aps,prl,showpacs,superscriptaddress,twocolumn,nofootinbib,floatfix,preprintnumbers,letterpaper,altaffilletter]{revtex4}
\usepackage{graphicx}
\usepackage{hyperref}
\usepackage{amssymb}
\usepackage{amsmath}
\usepackage{longtable}
\usepackage{rotating}
\usepackage{color}
\usepackage{fancyhdr}

\begin{document}


\title{
Upper Limits on a Stochastic Background of Gravitational Waves }


%
%
%

\newcommand*{\AG}{Albert-Einstein-Institut, Max-Planck-Institut f\"ur Gravitationsphysik, D-14476 Golm, Germany}
\affiliation{\AG}
\newcommand*{\AH}{Albert-Einstein-Institut, Max-Planck-Institut f\"ur Gravitationsphysik, D-30167 Hannover, Germany}
\affiliation{\AH}
\newcommand*{\AN}{Australian National University, Canberra, 0200, Australia}
\affiliation{\AN}
\newcommand*{\CH}{California Institute of Technology, Pasadena, CA  91125, USA}
\affiliation{\CH}
\newcommand*{\DO}{California State University Dominguez Hills, Carson, CA  90747, USA}
\affiliation{\DO}
\newcommand*{\CA}{Caltech-CaRT, Pasadena, CA  91125, USA}
\affiliation{\CA}
\newcommand*{\CU}{Cardiff University, Cardiff, CF2 3YB, United Kingdom}
\affiliation{\CU}
\newcommand*{\CL}{Carleton College, Northfield, MN  55057, USA}
\affiliation{\CL}
\newcommand*{\CO}{Columbia University, New York, NY  10027, USA}
\affiliation{\CO}
\newcommand*{\HC}{Hobart and William Smith Colleges, Geneva, NY  14456, USA}
\affiliation{\HC}
\newcommand*{\IU}{Inter-University Centre for Astronomy  and Astrophysics, Pune - 411007, India}
\affiliation{\IU}
\newcommand*{\CT}{LIGO - California Institute of Technology, Pasadena, CA  91125, USA}
\affiliation{\CT}
\newcommand*{\LM}{LIGO - Massachusetts Institute of Technology, Cambridge, MA 02139, USA}
\affiliation{\LM}
\newcommand*{\LO}{LIGO Hanford Observatory, Richland, WA  99352, USA}
\affiliation{\LO}
\newcommand*{\LV}{LIGO Livingston Observatory, Livingston, LA  70754, USA}
\affiliation{\LV}
\newcommand*{\LU}{Louisiana State University, Baton Rouge, LA  70803, USA}
\affiliation{\LU}
\newcommand*{\LE}{Louisiana Tech University, Ruston, LA  71272, USA}
\affiliation{\LE}
\newcommand*{\LL}{Loyola University, New Orleans, LA 70118, USA}
\affiliation{\LL}
\newcommand*{\MS}{Moscow State University, Moscow, 119992, Russia}
\affiliation{\MS}
\newcommand*{\ND}{NASA/Goddard Space Flight Center, Greenbelt, MD  20771, USA}
\affiliation{\ND}
\newcommand*{\NA}{National Astronomical Observatory of Japan, Tokyo  181-8588, Japan}
\affiliation{\NA}
\newcommand*{\NO}{Northwestern University, Evanston, IL  60208, USA}
\affiliation{\NO}
\newcommand*{\SE}{Southeastern Louisiana University, Hammond, LA  70402, USA}
\affiliation{\SE}
\newcommand*{\SA}{Stanford University, Stanford, CA  94305, USA}
\affiliation{\SA}
\newcommand*{\SR}{Syracuse University, Syracuse, NY  13244, USA}
\affiliation{\SR}
\newcommand*{\PU}{The Pennsylvania State University, University Park, PA  16802, USA}
\affiliation{\PU}
\newcommand*{\TC}{The University of Texas at Brownsville and Texas Southmost College, Brownsville, TX  78520, USA}
\affiliation{\TC}
\newcommand*{\TR}{Trinity University, San Antonio, TX  78212, USA}
\affiliation{\TR}
\newcommand*{\HU}{Universit{\"a}t Hannover, D-30167 Hannover, Germany}
\affiliation{\HU}
\newcommand*{\BB}{Universitat de les Illes Balears, E-07122 Palma de Mallorca, Spain}
\affiliation{\BB}
\newcommand*{\BR}{University of Birmingham, Birmingham, B15 2TT, United Kingdom}
\affiliation{\BR}
\newcommand*{\FA}{University of Florida, Gainesville, FL  32611, USA}
\affiliation{\FA}
\newcommand*{\GU}{University of Glasgow, Glasgow, G12 8QQ, United Kingdom}
\affiliation{\GU}
\newcommand*{\MU}{University of Michigan, Ann Arbor, MI  48109, USA}
\affiliation{\MU}
\newcommand*{\OU}{University of Oregon, Eugene, OR  97403, USA}
\affiliation{\OU}
\newcommand*{\RO}{University of Rochester, Rochester, NY  14627, USA}
\affiliation{\RO}
\newcommand*{\UW}{University of Wisconsin-Milwaukee, Milwaukee, WI  53201, USA}
\affiliation{\UW}
\newcommand*{\VC}{Vassar College, Poughkeepsie, NY 12604}
\affiliation{\VC}
\newcommand*{\WU}{Washington State University, Pullman, WA 99164, USA}
\affiliation{\WU}

\author{B.~Abbott}    \affiliation{\CT}
\author{R.~Abbott}    \affiliation{\CT}
\author{R.~Adhikari}    \affiliation{\CT}
\author{J.~Agresti}    \affiliation{\CT}
\author{B.~Allen}    \affiliation{\UW}
\author{J.~Allen}    \affiliation{\LM}
\author{R.~Amin}    \affiliation{\LU}
\author{S.~B.~Anderson}    \affiliation{\CT}
\author{W.~G.~Anderson}    \affiliation{\UW}
\author{M.~Araya}    \affiliation{\CT}
% ...

 \collaboration{The LIGO Scientific Collaboration, http://www.ligo.org}
 \noaffiliation
%
%

\date{\today}

\begin{abstract}
The Laser Interferometer Gravitational Wave Observatory (LIGO)
has performed a third science run with much improved sensitivities
of all three interferometers. We present an analysis of
approximately 200 hours of data acquired during this run, used to
search for a stochastic background of gravitational radiation. We
place upper bounds on the energy density stored as gravitational
radiation for three different spectral power laws. For the flat
spectrum, our limit of $\Omega_0 < 8.4\times 10^{-4}$ in the
69-156~Hz band is $\sim\!\!10^5$ times lower than the previous
result in this frequency range.

\end{abstract}

 \preprint{LIGO-P050003-E-R}

\maketitle

A stochastic background of gravitational waves could result from
the random superposition of an extremely large number of
unresolved and independent gravitational-wave (GW) emission
events. Such a background is analogous to the cosmic microwave
background radiation (CMBR), though its spectrum is unlikely to be
thermal. The emission events could be the result of cosmological
processes, as with the CMBR, but occurring much earlier after the
big bang---e.g., during inflation. The events could also be due to
more recent astrophysical processes. Placing upper limits on or
detecting the energy density of a stochastic background of
gravitational waves is one of the long term goals of GW detectors.

The stochastic background spectrum is typically characterized in
terms of a dimensionless quantity $\Omega_{\rm gw}(f)$: the GW
energy density per unit logarithmic frequency, divided by the
critical energy density $\rho_c$ to close the universe. The
critical density, and thus $\Omega_{\rm gw}(f)$, depend on the
Hubble expansion rate $H_0$; in this letter, all bounds on
$\Omega_{\rm gw}(f)$ will be for $H_0 =
72$~km~sec$^{-1}$~Mpc$^{-1}$ \cite{wmap}. We search for power laws
of the form $\Omega_{\rm gw}(f) = \Omega_\alpha (f/{\rm
100~Hz})^{\alpha}$. The choices of $\alpha$ are motivated by
potential source models \cite{maggiore}: ($\alpha=0$) predicted by
inflationary or cosmic string models; ($\alpha=2$) rotating
neutron stars; ($\alpha=3$) pre-big-bang cosmology.

Previous direct measurements of a
stochastic background, in the $\sim\!\!10$~Hz to $\sim\!\!10^4$~Hz
band accessible to earth-based detectors, have been limited to
establishing upper limits on $\Omega_{\rm gw}(f)$ greater than
unity, with the best and most recent result using LIGO's first
science data finding $\Omega_0 < 44$ \cite{s1stoch}. At much
lower frequencies, spacecraft Doppler tracking has established
$\Omega_0 < 0.027$ in the band $10^{-6}-10^{-3}$~Hz
\cite{cassini}, and radio pulsar timing has bounded $\Omega_0 <
10^{-7}$ in a decade band around $10^{-8}$~Hz \cite{mchugh}. In
this paper we report new limits on a stochastic GW background for
frequencies around $100$~Hz, using data from the LIGO GW
detectors. In terms of GW energy density, these limits are nearly
five orders of magnitude below previous limits in this frequency
band.

\begin{figure}[htbp!]
  \includegraphics[width=3.5in,angle=0]{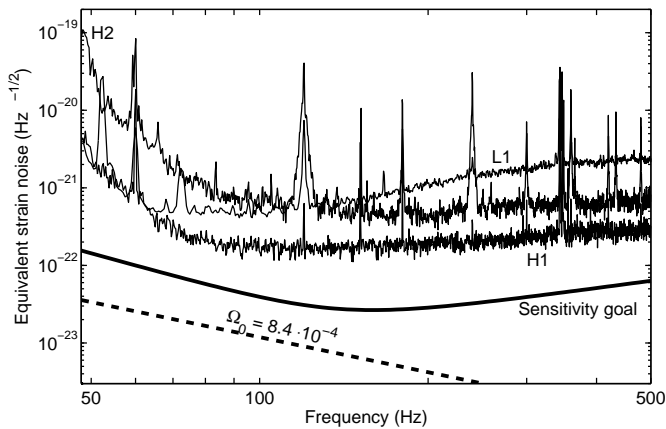}
  \caption{Reference sensitivity curves during the S3 data run, in
  terms of equivalent strain noise density. Also shown is the $f^{-3/2}$
  strain noise level corresponding to the upper limit found in this
  analysis, $\Omega_0 = 8.4 \times 10^{-4}$, and the strain noise
  goal for the two 4-km interferometers.}
  \label{fig:s3noise}
\end{figure}

LIGO is composed of three GW interferometers at two sites: the
4~km H1 and 2~km H2 detectors, collocated at Hanford, WA; and the
4~km L1 detector, located in Livingston Parish, LA
\cite{ligoproject}. Each detector is a power-recycled Michelson
interferometer, with 4~km (or 2~km) long Fabry-Perot cavities in
each of its orthogonal arms. These interferometers are sensitive
to quadrupolar oscillations in the space-time metric due to a
passing GW, measuring directly the GW strain amplitude. While the
detectors are still being commissioned to perform at their
designed sensitivity, several dedicated data collection runs have
been performed. The detector configuration and performance during
LIGO's first science run (S1) is described in Ref.~\cite{s1ligo}.
In this analysis we use data from the third science run (S3),
carried out from 31 Oct 2003--9 Jan 2004, with significantly
improved sensitivity compared to previous runs.
Fig.~\ref{fig:s3noise} shows reference amplitude spectra of
equivalent strain noise for S3. These were typical noise levels
for the detectors, though there was also significant variation
during the run; e.g., H1 displayed a general trend of improved
performance over time, with a minimum noise level at the end of
the run approximately a factor of 2 below that in
Fig.~\ref{fig:s3noise}.

To search for a stochastic background, we cross-correlate the
strain data from a pair of detectors, taking advantage of the fact
that the instrumental noise of one interferometer will, in
general, be uncorrelated with that of the other interferometers.
This is more clearly the case for the widely separated
interferometer pairs (i.e., L1 paired with H1 or H2), for which
there are only a few paths through which instrumental correlations
could arise. The H1-H2 cross-correlation, on the other hand, is
susceptible to a much broader range of potentially correlated
instrumental noise sources. This letter presents an analysis of
the inter-site pairs only; the H1-H2 measurement, though offering
potentially higher sensitivity due to their collocation, requires
additional techniques to address instrumental noise, and may be
presented in a later publication.

The cross-correlation is performed in the frequency domain, using
a linear filter that optimizes the expected signal-to-noise ratio,
given the detectors' noise spectra and the targeted stochastic
background (see Ref.~\cite{s1stoch} and references cited within).
Specifically, with $\tilde s_1(f)$ and
$\tilde s_2(f)$ representing the Fourier transforms of the strain
outputs of two detectors, the cross-correlation is computed as:
\begin{equation}\label{e:xcorr}
    Y = \int_{-\infty}^{\infty}df\ \int_{-\infty}^{\infty} df'\
        \delta_T(f-f')\,\tilde s_1^*(f)\,\tilde Q(f')\,\tilde s_2(f')\ ,
\end{equation}
where $\delta_T$ is a finite-time approximation to the Dirac delta
function. The optimal filter $\widetilde Q$ has the form:
\begin{equation}\label{e:optfilt}
    \widetilde Q(f) = {\cal N} \frac{\gamma(f) S_{\rm gw}(f)}
                 {P_1(f) P_2(f)}\ ,
\end{equation}
where $\cal N$ is a normalization factor, $P_1$ and $P_2$ are the
strain noise power spectra of the two detectors, $S_{\rm gw}$ is
the strain power spectrum of the stochastic background being
searched for
($S_{\rm gw}(f) = (3H_0^2/10 \pi^2)f^{-3} \Omega_{\rm gw}(f)$),
and the factor $\gamma$ is called the
overlap reduction function \cite{flan}. This factor, defined so
that its absolute value is at most unity, gives the frequency
variation of the cross-correlation arising from an isotropic
stochastic background, for separated or non-aligned detectors
($\gamma(f)=1$ at all frequencies for the co-located detector pair,
H1 and H2).

The optimal filter is derived assuming that the intrinsic detector
noise is Gaussian and stationary over the measurement time,
uncorrelated between detectors, and uncorrelated with and much
greater in power than the stochastic GW signal. Under these
assumptions the expected variance, $\sigma^2_Y$, of the
cross-correlation is dominated by the noise in the individual
detectors, whereas the expected value of the cross-correlation $Y$
depends on the stochastic background power spectrum:
\begin{gather}\label{e:sigma}
    \sigma_Y^2 \equiv \langle Y^2 \rangle - \langle Y \rangle^2 \approx
               \frac{T}{2} \int_{0}^{\infty} df\,P_1(f)\,|\widetilde Q(f)|^2
               \,P_2(f) \\
\label{e:mu}
    \langle Y \rangle = T \int_{0}^{\infty} df\, \gamma(f)
    S_{\rm gw}(f) \widetilde Q(f)\ ,
\end{gather}
where $T$ is the duration of the measurement.

\emph{Analysis Details.}--The analysis is implemented similarly to
the method detailed in Ref.~\cite{s1stoch}. The data set from a
given interferometer pair is divided into equal-length intervals,
and the cross-correlation $Y$ and theoretical $\sigma_Y$ are
calculated for each interval, yielding a set $\{Y_I,
\sigma_{Y_I}\}$ of such values, with $I$ labelling the intervals.
This data segmentation is useful for dealing with long-term
non-stationarity of the detector noise, by choosing an interval
length over which the noise is relatively stationary. The interval
length for this analysis is 60~sec. The cross-correlation values
are combined to produce a final cross-correlation estimator,
$Y_{\rm opt}$, that maximizes the signal-to-noise ratio, and has
variance $\sigma_{\rm opt}^2$:
\begin{equation}\label{e:Yopt}
  \begin{array}{cc}
    Y_{\rm opt} = \sum_I \sigma_{Y_I}^{-2} Y_I / \sigma_{\rm
    opt}^{-2}\ ,
    &
    \:\sigma_{\rm opt}^{-2} = \sum_I \sigma_{Y_I}^{-2}\ .
  \end{array}
\end{equation}
The normalization factor $\cal N$ is defined such that the point
estimate of $\Omega_{\alpha}$ and its standard deviation are given by:
$\widehat\Omega_\alpha =
Y_{\rm opt}/T,\
\sigma_{\Omega_\alpha} = \sigma_{\rm opt}/T$.

Before computing the cross-correlation, each 60~sec data interval
is decimated (from 16384~Hz to 1024~Hz), high-pass filtered (40~Hz
cut-off) and Hann-windowed. The windowing step protects against
spectral leakage of strong lines that may be present in the data,
but at the same time a Hann window reduces the effective length of
the interval by nearly a factor of 2 (approximately the mean value
of the Hann window). To recover the loss in signal-to-noise, the
data intervals are overlapped by 50\%, so that each data point
receives full weighting in the analysis (except for end effects).
This introduces some correlation between the $Y_I$ and
$\sigma_{Y_I}$ for adjacent values of $I$, complicating the
formulae for the optimal estimator and its variance, derivations
of which may be found in Ref.~\cite{ovlpwin}.

The detectors' strain noise power spectral densities (PSDs) are
estimated for each interval in order to calculate the optimal
filter $\widetilde Q(f)$. Welch's modified periodogram method of
PSD estimation is used, averaging 58 periodograms formed from
4~sec-long, 50\% overlapping data windows. The PSD for interval
$I$ is formed from the two 60~sec data intervals preceding and
following, but excluding the data within, interval $I$. This
technique eliminates a bias (underestimate) in the
cross-correlation that would otherwise exist, due to non-zero
covariance between the $\tilde s_1^* \cdot \tilde s_2$
cross-spectrum and the corresponding power spectra. However,
short-term changes (typically increases) in detector noise may
produce outliers, because excess instrumental noise within
interval $I$ is not reflected in its PSD. This is addressed by
applying a consistency test on $\sigma_{Y_I}$ over 3 consecutive
intervals; those intervals for which the $\sigma_{Y_I}$ differ by
more than $20\%$ are eliminated from the analysis. For this
analysis, approximately $20\%$ of the data were rejected by this
cut, leaving 218 hours.

%
\begin{figure}[htbp!]
  \includegraphics[width=3.5in,angle=0]{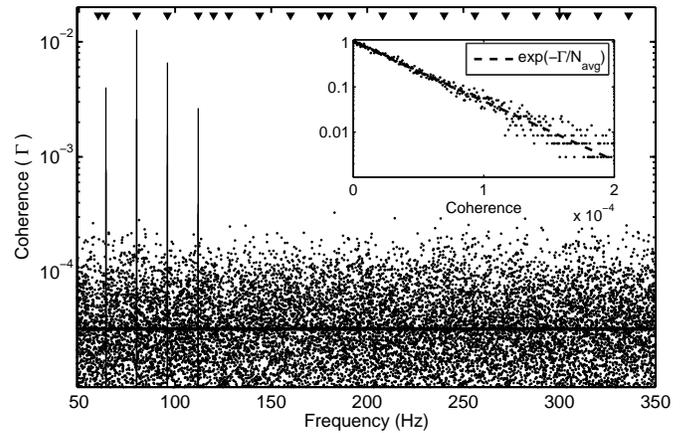}
  \caption{Coherence between H1 and L1 during S3, showing a few small, but
  significant, coherent peaks at multiples of 16~Hz. The horizontal line
  corresponds
  to the statistical expectation value of $1/N_{\rm avg} = 3.3 \times 10^{-5}$,
  where $N_{\rm avg}$ is the number of periodogram averages. 
  The inverted triangles
  at the top of the graph indicate the discrete frequencies omitted from the
  analysis. The inset histogram shows that the coherence values ($\Gamma$)
  follow the expected exponential distribution (dashed line).}
  \label{fig:h1l1coh}
\end{figure}
%

\begin{table*}
\begin{ruledtabular}
\begin{tabular}{cccccccc}

Power & Frequency & Pt Estimate & Statistical Error &
\multicolumn{2}{c}{Calibration Error} &
\multicolumn{2}{c}{Upper Limits}
\\ 
\cline{7-8}
Law & Range & $\widehat\Omega_\alpha$ & $\sigma_{\Omega_\alpha}$ & 
(H1) & (L1) & 
$\Omega_{\rm gw}(f)$ &
$S_{\rm gw}^{1/2}(f)\ ({\rm Hz^{-1/2}})$ \\

\hline

$\alpha=0$ & $69-156$~Hz & $-6.0\times 10^{-4}$ & 
$7.0\times 10^{-4}$ & $\pm 9\%$ & $\pm 15\%$ &
$8.4\times 10^{-4}$ &
$1.2\times 10^{-23}\,(f/100~{\rm Hz})^{-3/2}$ \\

$\alpha=2$ & $73-244$~Hz & $-4.7\times 10^{-4}$ &
$7.2\times 10^{-4}$ & $\pm 9\%$ & $\pm 15\%$ &
$9.4\times 10^{-4}\,(f/100~{\rm Hz})^2$ & 
$1.2\times 10^{-23}\,(f/100~{\rm Hz})^{-1/2}$ \\

$\alpha=3$ & $76-329$~Hz & $-4.0 \times 10^{-4}$ &
$6.2\times 10^{-4}$ & $\pm 9\%$ & $\pm 15\%$ &
$8.1\times 10^{-4}\,(f/100~{\rm Hz})^3$ & 
$1.2\times 10^{-23}$ \\

%






\end{tabular}
\end{ruledtabular}
\caption{Results of the cross-correlation of LIGO's H1 and L1
interferometers, analyzed for a potential power-law stochastic
background of the form: $\Omega_{\rm gw}(f) = \Omega_\alpha (f/100~{\rm
Hz})^{\alpha}$. The frequency range for each $\alpha$ is the band
that contributes 99\% of the full sensitivity, as determined by
the inverse variance. All results correspond to the specified
band, and an observation time of 218~hr. 
90\%-confidence Bayesian upper limits on $\Omega_{\rm gw}(f)$ 
(also expressed as limits on the strain noise density
$S_{\rm gw}^{1/2}(f)$) are calculated from the point estimates 
and statistical errors, marginalising over a $\pm9\%$ and
$\pm15\%$ uncertainty in the calibration magnitude of the H1 and
L1 detectors.}
\label{tab:results}
\end{table*}

To compute the cross-correlation (Eq.~\ref{e:xcorr}), the raw
detector data are calibrated, in the frequency domain, into strain
units using interferometer response functions. These functions are
calculated once per 60~sec, using a measurement of the response of
an interferometer to a sinusoidal calibration force, averaged over
60~sec. The frequency domain values of $\{\tilde s_1, \tilde
s_2\}$, given at a frequency spacing of $1/60$~Hz, are binned to
the resolution of the optimal filter (frequency spacing of
$1/4$~Hz), and the integrations in Eqs.~\ref{e:xcorr} and
\ref{e:sigma} are performed for the different $S_{\rm gw}(f)$.

As was important in the earlier analysis of Ref.~\cite{s1stoch},
frequency bins corresponding to known or potential instrumental
correlation artifacts are excluded from the frequency domain
integrations. An obvious example of inter-site correlations comes
from the 60-Hz AC supply lines used to power the detectors. The
60-Hz modulation and its harmonics are present to some degree in
the detector electronics, and thereby infiltrate the strain output
signal (as Fig.~\ref{fig:s3noise} shows). Between L1 and H1,2, the
power lines tend to be well correlated over time scales shorter
than $\sim\!\! 100$~sec, with ever-decreasing correlation over
longer times. To exclude the possibility of any residual long-term
power line correlation, the (60~Hz, 120~Hz, \ldots) bins are
excluded from the integration.

Another narrow-band source of instrumental correlation stems from
imperfections in the detectors' data acquisition systems. The
peaks in Fig.~\ref{fig:h1l1coh}, at multiples of 16~Hz, were
produced by slight but periodic corruption of the data at each
site. The data acquisition timing at each site is controlled by
clocks synchronized to the 1 pulse-per-second signals produced by
Global Positioning System (GPS) receivers. The 16~Hz periodicity
of the data corruption was controlled by these clocks, resulting
in persistent inter-site correlations at multiples of 16~Hz. Thus,
the (16~Hz, 32~Hz, \ldots) bins are also excluded from the
integration. (After S3, the offending clock modules were
identified and the problem was corrected.)

\emph{Results.}--For each power law searched for, we calculate the
optimal cross-correlation statistic and its variance (Eqs.~\ref{e:Yopt}).
The H1-L1 results are summarized in Table~\ref{tab:results}.
The H2-L1 correlation is approximately a
factor of 7 less sensitive than H1-L1, due to the higher noise
level of H2; the H2-L1 results are thus not used for the upper
limits results (and are not shown), but they are consistent,
within their error bars, with the H1-L1 results.

Systematic errors due to unresolved time variations in the
interferometers' calibration and power spectra were investigated
and determined to be small compared to the statistical error
$\sigma_{\Omega_\alpha}$. Phase calibration uncertainties and timing
errors are also negligible. Not negligible are interferometer
amplitude calibration uncertainties, estimated as a $\pm 9\%$
($\pm 15\%$) frequency-independent uncertainty in the strain response
magnitude for H1 (L1).

We construct a Bayesian posterior probability distribution for 
$\Omega_{\alpha}$ using the optimal point estimate 
$\widehat\Omega_{\alpha}$ and statistical error 
$\sigma_{\Omega_\alpha}$, marginalising over the unknown calibration 
magnitudes (see, e.g., \cite{loredo-sn1987a}).
The prior probability distribution for $\Omega_{\alpha}$ is taken 
to be uniform from 0 to 0.02 (the maximum value corresponding to 
the largest background that is still consistent with the lowest single
interferometer strain noise); the prior distributions for the 
calibration magnitudes are taken to be uniform between $1\pm0.09$ 
(for H1) and $1\pm0.15$ (for L1).
The 90\% probability upper limit is then that value of 
$\Omega_{\alpha}$ for which 90\% of the posterior distribution 
lies between 0 and the upper limit.
(The upper limit is relatively insensitive to reasonable changes 
in the maximum value of the prior distribution for $\Omega_{\alpha}$,
as well as to reasonable changes in the prior distributions for the 
calibration magnitudes.)

\begin{figure}[htbp!]
  \includegraphics[width=3.5in,angle=0]{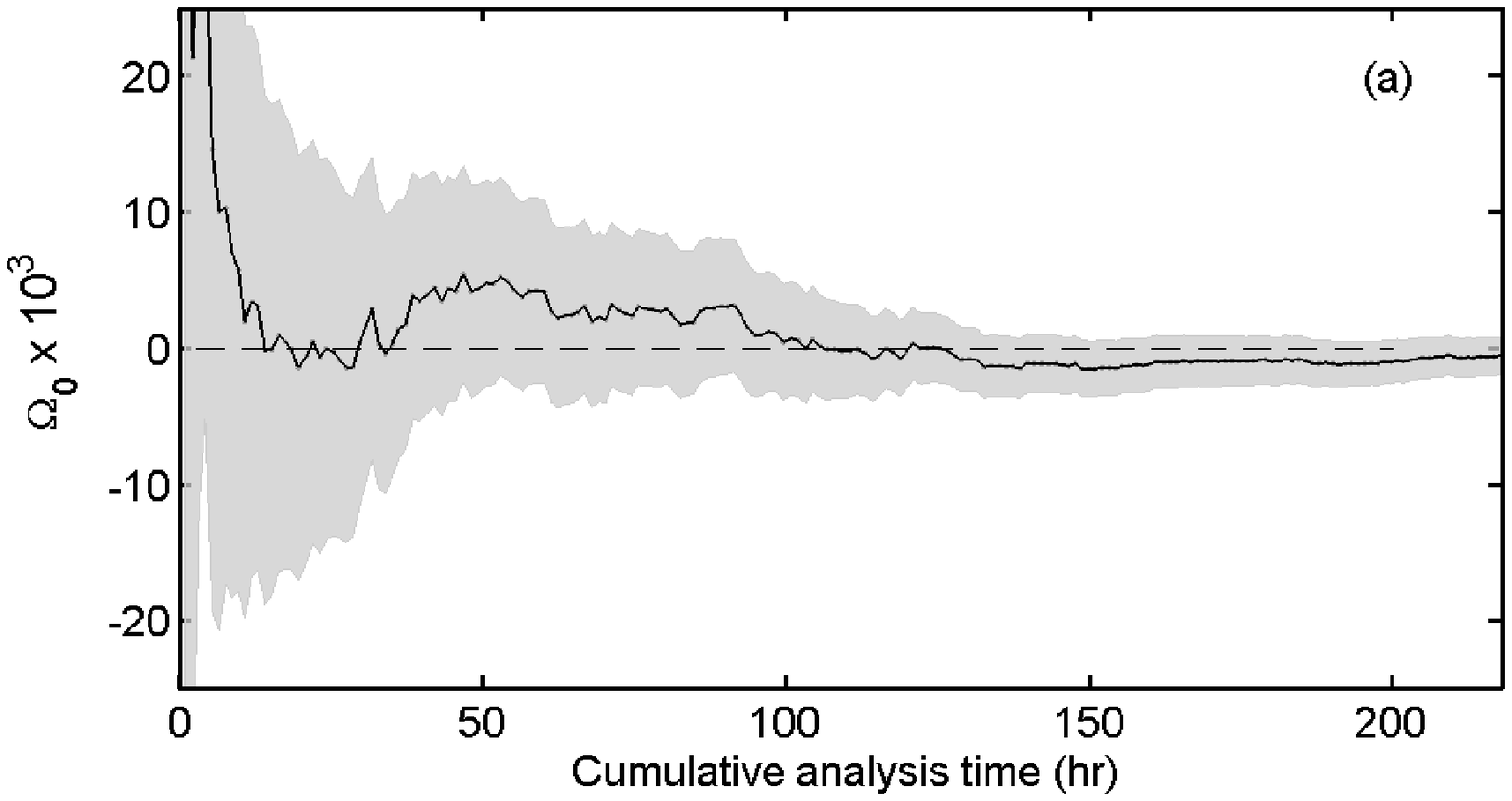}
  \includegraphics[width=3.5in,angle=0]{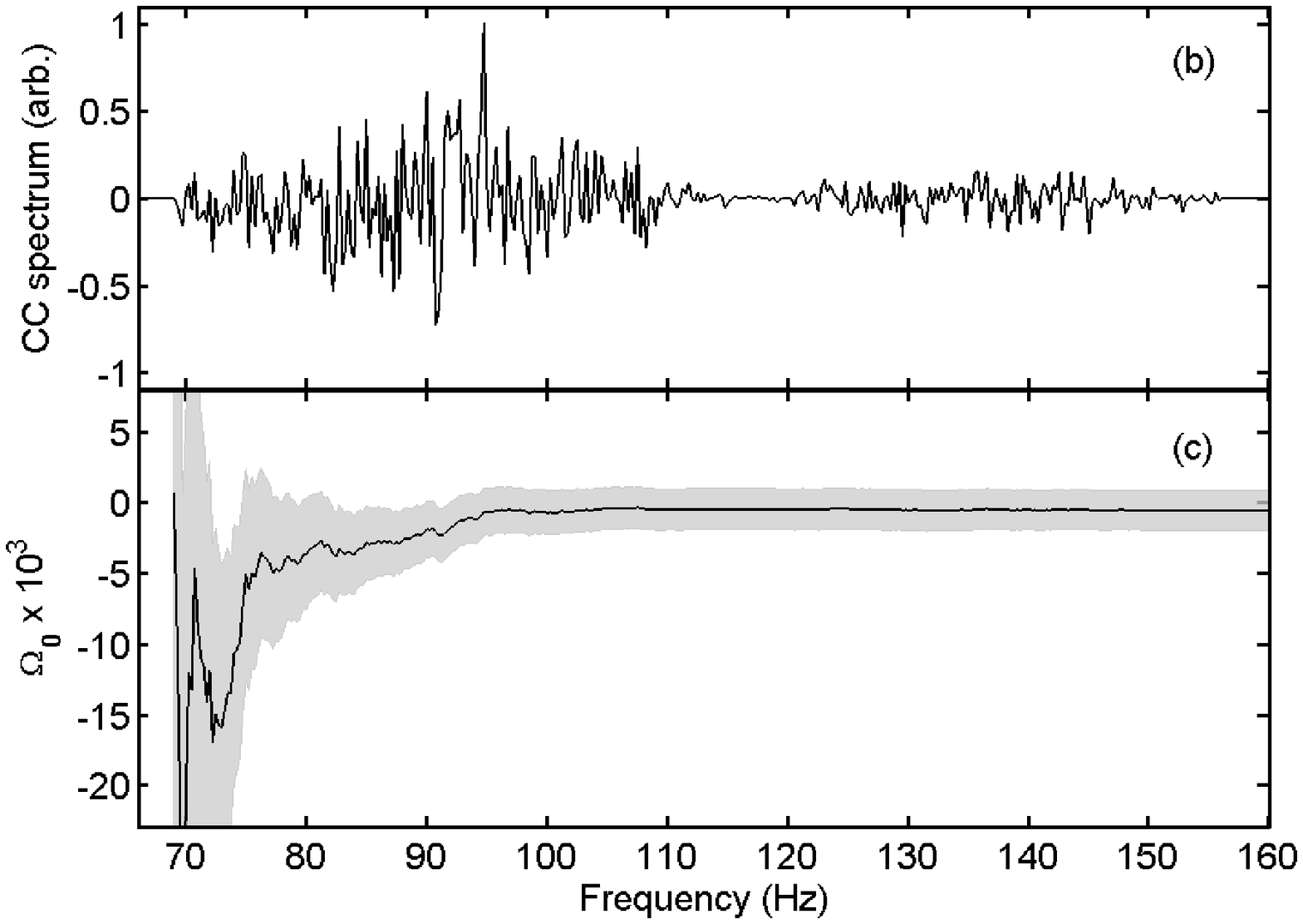}
  \caption{The estimate $\widehat \Omega_{\alpha}$, for H1-L1 and
  $\alpha = 0$: 
  (a) As a function of the amount of data analyzed. The shaded
  region shows the $\pm 2\sigma_{\Omega_0}$ band on $\widehat \Omega_{0}$.
  (b) The real part of the H1-L1 cross-correlation spectrum, in arbitrary
  units. 
  (c) As a function of the frequency range analyzed.
  The shaded region shows the $\pm 2\sigma_{\Omega_0}$ band on 
  $\widehat \Omega_{0}$, cumulative in frequency from $69~$Hz to $f$.}
  \label{fig:h1l1data}
\end{figure}

The estimates for $\Omega_{\alpha}$ are entirely consistent
with no stochastic background, within the sensitivity of the
measurement. Furthermore, the cross-correlation spectrum (i.e.,
the integrand of Eq.~\ref{e:xcorr}) shows no distinct features
(see Fig.~\ref{fig:h1l1data}), and the $\approx 26,000$ values of
$Y_I$, with mean removed and normalized by the $\sigma_{Y_I}$,
follow the expected normal distribution.

\emph{Conclusions.}--The energy density in a primordial background
of GWs is constrained by big-bang nucleosynthesis theory, giving a
conservative bound of: $\int d(\ln f) \Omega_{\rm gw}(f) < 1.1
\times 10^{-5}$ \cite{maggiore}; if all the GW energy were
concentrated in our sensitive band, this is still $60\times$ below
the limit set here. A background from astrophysical sources would
be generated at much later cosmic times, and thus not subject to
the above bound. In the LIGO band, such a background could be
generated by the superposition of many short-lived signals, such
as supernovae bursts or quasi-normal modes from compact objects
(in the kHz region) and rotating neutron stars (in the tens to
hundreds of Hz band). Uncertainties in the theoretical models are
large, though the most optimistic predictions peak at $\Omega_{\rm
gw}(f) \sim 10^{-7}$\cite{maggiore}. Nonetheless, the results
presented here provide direct, measured limits to a stochastic
background, that, in terms of energy density, are nearly $10^5$
lower than previous measurements. Eventually, with 1-year of data
at design sensitivity, the LIGO detectors will be sensitive at a
level several times below the nucleosynthesis bound.

\begin{acknowledgments}
The authors gratefully acknowledge the support of the United States National 
Science Foundation for the construction and operation of the LIGO Laboratory 
and the Particle Physics and Astronomy Research Council of the United Kingdom, 
the Max-Planck-Society and the State of Niedersachsen/Germany for support of 
the construction and operation of the GEO600 detector. The authors also 
gratefully acknowledge the support of the research by these agencies and by the 
Australian Research Council, the Natural Sciences and Engineering Research 
Council of Canada, the Council of Scientific and Industrial Research of India, 
the Department of Science and Technology of India, the Spanish Ministerio de 
Educacion y Ciencia, the John Simon Guggenheim Foundation, the Leverhulme Trust,
the David and Lucile Packard Foundation, the Research Corporation, and the Alfred P. Sloan Foundation. 
\end{acknowledgments}


\end{document}